\documentclass[twocolumn,showpacs,preprintnumbers,amsmath,amssymb]{revtex4}

\usepackage{graphicx}
\usepackage{dcolumn}
\usepackage{bm}

\begin{document}

\title{General model for Apollonian networks}
\author{Zhongzhi Zhang}
\email{xinjizzz@sina.com}
\author{Lili Rong}
\email{llrong@dlut.edu.cn}
\affiliation {%
Institute of Systems
Engineering, Dalian University of Technology,
2 Ling Gong Rd., Dalian 116023, Liaoning, China}%

\date{\today}

\begin{abstract}
We introduce a general deterministic model for Apollonian Networks
in an iterative fashion. The networks have small-world effect and
scale-free topology. We calculate the exact results for the degree
exponent, the clustering coefficient and the diameter. The major
points of our results indicate that (a) the degree exponent can be
adjusted in a wide range, (b) the clustering coefficient of each
individual vertex is inversely proportional to its degree and the
average clustering coefficient of all vertices approaches to a
nonzero value in the infinite network order, and (c) the diameter
grows logarithmically with the number of network vertices.
\end{abstract}



\maketitle


\section{Introduction}

In the past few years we have witnessed an upsurge in the research
of a wide range of complex systems which can be described in terms
of networks------vertices connected together by edges
\cite{AlBa02,DoMe03,SaVe04,Ne03}. Various empirical studies have
shown that most real-life systems exhibit the following three
properties: power-law degree distribution \cite{BaAl99}, high
clustering coefficient \cite{WaSt98} and small average path length
(APL) which implies that their average intervertex distances grow
logarithmically with the number of vertices or slower.

In order to mimic real-world systems, a wide variety of models have
been proposed~\cite{AlBa02,DoMe03,SaVe04,Ne03}, most of which are
random. Randomness may be in line with the major features of
real-life networks, but as mentioned by Barab\'asi et al., it makes
harder to gain a visual understanding of how networks are shaped,
and how do different vertices relate to each other~\cite{BaRaVi01}.
Therefore, it would be of major theoretical interest to construct
deterministic models having similar structural characteristics as
systems in nature and society. A strong advantage of deterministic
networks is that it is often possible to compute analytically their
properties, which may be compared with empirical data from real-life
systems. In the last few years, deterministic networks have been
intensively
studied~\cite{BaRaVi01,CoOzPe00,CoSa02,DoGoMe02,JuKiKa02,CoFeRa04,RaBa03,No03,ZhWaHuCh04,IgYa05,ZhRoGo05}.

Recently, In relation to the problem of Apollonian packing, Andrade
\emph{et al.} introduced Apollonian networks~\cite{AnHeAnSi05} which
were also proposed by Doye and Massen in~\cite{DoMa05}. Apollonian
networks are part of a deterministic growing type of networks and
have received much attention. Except their further properties
\cite{ZhCoFeRo05}, stochastic versions \cite{ZhYaWa05,ZhRoCo05} and
potential implications \cite{DoMa05,DoMa05b}, many peculiar results
about some famous models upon Apollonian networks have been
found~\cite{AnHeAnSi05,AnHe05,LiGaHe04,AnMi05}.

In this paper, we do an extensive study on Apollonian networks. In
an iterative way we propose a general model for Apollonian networks
by including a parameter. Apollonian
networks~\cite{AnHeAnSi05,DoMa05,ZhCoFeRo05} are special cases of
the present model. The deterministic construction of our model
enables one to obtain the analytic solution for the degree
distribution, clustering coefficient and the diameter. Moreover, by
tuning the parameter, we can obtain scale-free networks with a
variety of degree exponents and clustering coefficient.

\section{Brief introduction to Apollonian networks}
\begin{figure}
\begin{center}
\includegraphics[width=0.55\textwidth]{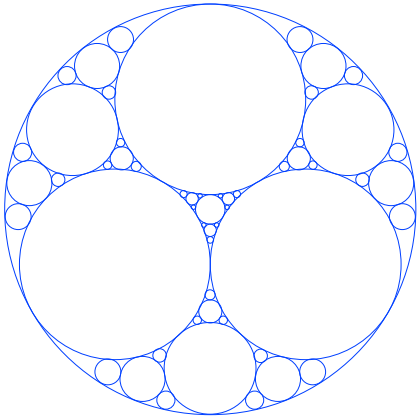} \
\end{center}
\caption[kurzform]{\label{Fig1} (Color online) A two-dimensional
Apollonian packing of disks. }
\end{figure}

%
To define Apollonian networks, we first introduce Apollonian packing
(see Fig. \ref{Fig1} for the case of two dimension). The classic
two-dimensional Apollonian packing is constructed as follows.
Initially three mutually touching disks are inscribed inside a
circular space which is to be filled. The interstices of the initial
disks and circle are curvilinear triangle to be filled. This initial
configuration is called generation $t=0$. Then in the first
generation $t=1$, four disks are inscribed, each touching all the
sides of the corresponding curvilinear triangle. For next
generations we simply continue by setting disks in the newly
generated curvilinear triangles. The process is repeated
indefinitely for all the new curvilinear triangles. In the limit of
infinite generations we obtain the Apollonian packing, in which the
circular space is completely filled with disks of various sizes. The
translation from Apollonian packing construction to Apollonian
network generation is quite straightforward: vertices of the network
represent disks and two vertices are connected if the corresponding
disks are tangent~\cite{AnHeAnSi05,DoMa05}.

The two-dimensional Apollonian network can be easily generalized to
high-dimensions ($d$-dimensional, $d\geq 2$)
\cite{DoMa05,ZhCoFeRo05} associated with other self-similar packings
\cite{MaHeRi04}. The $d$-dimensional Apollonian packings start with
$d+1$ mutually touching $d$-dimensional hyperspheres that is
enclosed within and touching a larger $d$-dimensional hyperspheres,
with $d+2$ curvilinear $d$-dimensional simplex ($d$-simplex) as
their interstices, which are to be filled in successive generations.
If each $d$-hypersphere corresponds to a vertex and vertices are
connected if the corresponding $d$-hyperspheres are in contact, then
$d$-dimensional Apollonian networks are gotten.

According to the construction process of $d$-dimensional Apollonian
packings, in Ref.~\cite{ZhCoFeRo05}, an generation algorithm for
$d$-dimensional Apollonian networks was proposed and further
properties of the networks were investigated. The $d$-dimensional
Apollonian networks are scale-free and display small-world
effect~\cite{AnHeAnSi05,DoMa05,ZhCoFeRo05}. The degree exponent is
$\gamma = 1+\frac{\ln (d+1)}{\ln d}$. Both their diameter and
average path length grow logarithmically with the number of network
vertices. The clustering coefficient is large, for two- and
three-dimensional Apollonian networks, it approaches 0.8284 and
0.8852, respectively.

\section{Iterative algorithm for the general model of Apollonian networks}
%
Our model is constructed in an iterative fashion. Before introducing
the algorithm we give the following definitions on a graph
(network). The term \emph{size} refers to the number of edges in a
graph. The number of vertices in the graph is called its
\emph{order}. When two vertices of a graph are connected by an edge,
these vertices are said to be \emph{adjacent}, and the edge is said
to join them. A \emph{complete graph} is a graph in which all
vertices are adjacent to one another. Thus, in a complete graph,
every possible edge is present. The complete graph with $d$ vertices
is denoted as $K_d$ (also referred in the literature as
$d$-\emph{clique}; see \cite{We01}). Two graphs are
\emph{isomorphic} when the vertices of one can be relabeled to match
the vertices of the other in a way that preserves adjacency. So all
$d$-cliques are isomorphic to one another.

The general model for Apollonian networks after $t$ generations are
denoted by $A(d,t)$, $d\geq 2, t\geq 0$. Then at step $t$, the
considered networks are constructed as follows: For $t=0$, $A(d,0)$
is a complete graph $K_{d+2}$ (or $(d+2)$-clique). For $t\geq 1$,
$A(d,t)$ is obtained from $A(d,t-1)$. For each of the existing
subgraphs of $A(d,t-1)$ that is isomorphic to a $(d+1)$-clique and
created at step $t-1$, $m$ new vertices are created, and each is
connected to all the vertices of this subgraph. The growing process
is repeated until the network reaches the desired order. When $m=1$,
the networks reduce to the deterministic Apollonian
networks~\cite{AnHeAnSi05,DoMa05,ZhCoFeRo05}.

Now we compute the order and size of the networks. Let $n_v(t)$,
$n_e(t)$ and $K_{(d+1),t}$ be the numbers of vertices, edges and
$(d+1)$-cliques created at step $t$, respectively. Note that the
addition of each new vertex leads to $d+1$ new $(d+1)$-cliques and
$d+1$ new edges, so we have $n_v(t)=mK_{(d+1),t-1}$,
$n_e(t)=(d+1)n_v(t)$,  and $K_{(d+1),t}=(d+1)n_v(t)$. Thus one can
easily obtain $K_{(d+1),t}=m(d+1)K_{(d+1),t-1}=(d+2)[m(d+1)]^{t}$
($t\geq0$), $n_v(t)=(d+2)m^{t}(d+1)^{t-1}$ ($t>0$) and
$n_e(t)=(d+2)m^{t}(d+1)^{t}$ ($t>0$). So the number of network
vertices increases exponentially with time, which is similar to many
real-life networks such as the World Wide Web. From above results,
we can easily compute the size and order of the networks. The total
number of vertices $N_t$ and edges $E_t$ present at step $t$ is
\begin{eqnarray}\label{Nt}
N_t&=&\sum_{t_i=0}^{t}n_v(t_i)\nonumber\\
&=&\frac{m(d+2)[m^t(d+1)^{t}-1]}{m(d+1)-1}+d+2
\end{eqnarray}
and
\begin{eqnarray}\label{Et}
E_t
=\sum_{t_i=0}^{t}n_e(t_i)\nonumber\\
=\frac{m(d+2)(d+1)[m^t(d+1)^{t}-1]}{m(d+1)-1}+\frac{(d+2)(d+1)}{2}
\end{eqnarray}
respectively. So for large $t$, The average degree $\overline{k}_t=
\frac{2E_t}{N_t}$ is approximately $2(d+1)$.

\section{Topology properties of the networks}

\subsection{Degree distribution}
When a new vertex $i$ is added to the graph at step $t_i$, it has
degree $d+1$ and forms $d+1$ new $(d+1)$-cliques. Let
$K_{(d+1)}(i,t)$ be the number of newly-created $(d+1)$-cliques at
step $t$ with $i$ as one vertex of them, which will create new
vertices connected to the vertex $i$ at step $t+1$. At step $t_i$,
$K_{(d+1)}(i,t_i)=d+1$. From the iterative process, we can see that
each new neighbor of $i$ generates $d$ new $(d+1)$-cliques with $i$
as one vertex of them. Let $k_i(t)$ be the degree of $i$ at step
$t$. It is not difficult to find following relations:
\begin{equation}
\Delta k_i(t)=k_i(t)-k_i(t-1)=mK_{(d+1)}(i,t-1)\nonumber\\
\end{equation}
and
\begin{equation}
K_{(d+1)}(i,t)=dmK_{(d+1)}(i,t-1)=(d+1)(dm)^{t-t_{i}}\nonumber\\
\end{equation}
Then the degree of vertex $i$ becomes
\begin{eqnarray}\label{Ki}
k_i(t)&=&k_i(t_i)+m\sum_{\tau=t_i}^{t-1}
K_{(d+1)}(i,\tau)\nonumber\\&=&\frac{m(d+1)[(md)^{t-t_{i}}-1]+d^{2}-1}{d-1}
\end{eqnarray}

Since the degree of each vertex has been obtained explicitly as in
Eq.~(\ref{Ki}), we can get the degree distribution via its
cumulative distribution, i.e. $P_{cum}(k) \equiv \sum_{k^\prime \geq
k} N(k^\prime,t)/N_t \sim k^{1-\gamma}$, where $N(k^\prime,t)$
denotes the number of vertices with degree $k^\prime$. The analytic
computation details are given as follows. For a degree $k$
\begin{equation*}
\frac{m(d+1)[(md)^{t-p}-1]+d^{2}-1}{d-1}
\end{equation*}
there are  $n_v(p)=(d+2)m^{p}(d+1)^{p-1}$ vertices with this exact
degree, all of which were born at step $p$. All vertices with birth
time at $p$ or earlier have this and a higher degree. So we have
\begin{eqnarray}
\sum_{k' \geq k}
N(k',t)=\sum_{s=0}^{p}n_v(s)\nonumber\\
=\frac{m(d+2)[m^{p}(d+1)^{p}-1]}{m(d+1)-1}+d+2
\end{eqnarray}
As the total number of vertices at step $t$ is given
in Eq.~(\ref{Nt}) we have
\begin{eqnarray}
\left[\frac{m(d+1)[(md)^{t-p}-1]+d^{2}-1}{d-1}\right]^{1-\gamma}\nonumber\\
=\frac{\frac{m(d+2)[m^{p}(d+1)^{p}-1]}{m(d+1)-1}+d+2}{\frac{m(d+2)[m^t(d+1)^{t}-1]}{m(d+1)-1}+d+2}\nonumber
\end{eqnarray}
Therefore, for large $t$ we obtain
\begin{equation*}
\left[(md)^{t-p}\right]^{1-\gamma}=[m(d+1)]^{p-t}
\end{equation*}
and
\begin{equation}\label{Gamma}
\gamma \approx 1+\frac{\ln [m(d+1)]}{\ln(md)}
\end{equation}
For $m=1$, Eq. (\ref{Gamma}) recovers the results previously
obtained in Refs.~\cite{AnHeAnSi05,DoMa05,ZhCoFeRo05}.

\subsection{Clustering coefficient}
We can go beyond the degree distribution and obtain the analytical
expression for clustering coefficient $C(k)$ of an individual vertex
as a function of its degree $k$. By definition, clustering
coefficient~\cite{WaSt98} $C_{i}$ of a vertex $i$ is the ratio of
the total number $e_{i}$ of edges that actually exist between all
$k_{i}$ its nearest neighbors and the number $k_{i}(k_{i}-1)/2$ of
all possible edges between them, i.e.
$C_{i}=2e_{i}/[k_{i}(k_{i}-1)]$. The clustering coefficient of the
whole network is the average of $C_i's $ over all the vertices.

\begin{figure}
\begin{center}
\includegraphics[width=0.45\textwidth]{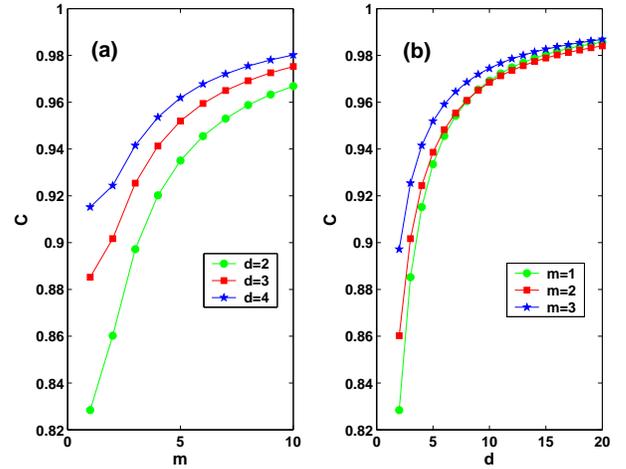} \
\end{center}
\caption[kurzform]{\label{Fig2} (Color online) The dependence
relation of $C$ on $d$ and $m$. }
\end{figure}

When a vertex is generated it makes connections to all the vertices
of a $(d+1)$-clique whose vertices are completely interconnected. It
follows that its degree and clustering coefficient are $d+1$ and 1,
respectively. In the following iterative steps, if its degree
increases one by a newly created vertex connecting to it, then there
must be $d$ existing neighbors of it attaching to the new vertex at
the same time. Thus for a vertex of degree $k$, we have
\begin{equation}\label{Ck}
C(k)= {{{d(d+1)\over 2}+ d(k-d-1)} \over {k(k-1)\over 2}}=
\frac{2d(k-\frac{d+1}{2})}{k(k-1)}
\end{equation}
where the last expression in Eq. (\ref{Ck}) is obtained after some
algebraic manipulations. From Eq. (\ref{Ck}), one can easily see
that $C(k)$ depends on degree $k$ and $d$. The asymptotic behavior
for large $k$ is $C(k)\sim 1/k$, which implies that $C(k)$ is
inversely proportional to the degree. Interestingly, we recover in
our model the same scaling behavior of $C(k)$ found in other
models~\cite{DoGoMe02,CoFeRa04,RaBa03,No03,IgYa05,ZhRoGo05,AnHeAnSi05,DoMa05,ZhCoFeRo05,ZhYaWa05,ZhRoCo05}
and some real-life systems~\cite{RaBa03}.

Using Eq. (\ref{Ck}), we can obtain the clustering $C$ of the
networks at step $t$:
\begin{equation}\label{AC}
C=\frac{1}{N_{t}}\sum_{r=0}^{t}
\frac{2d(D_r-\frac{d+1}{2})n_v(r)}{D_r(D_r-1)}
\end{equation}
where the sum is the total of clustering coefficient for all
vertices and $D_r=\frac{m(d+1)[(md)^{t-r}-1]+d^{2}-1}{d-1}$ shown by
Eq. (\ref{Ki}) is the degree of the vertices created at step $r$.

In the infinite network order limit ($N_{t}\rightarrow \infty$), Eq.
(\ref{AC}) converges to a nonzero value. When $d=2$, for $m=1$, 2
and 3, $C$ equal to 0.8284, 0.8602 and 0.8972, respectively. When
$m=2$, for $d=2$, 3 and 4, $C$ are 0.8602, 0.9017 and 0.9244,
respectively. Therefore, the clustering coefficient of our networks
is very high. Moreover, similarly to the degree exponent $\gamma$,
clustering coefficient $C$ is determined by $d$ and $m$. Fig.
\ref{Fig2} shows the dependence of $C$ on $d$ and $m$. From Fig.
\ref{Fig2} (a) and (b), one can see that for any fixed $m$, $C$
increases with $d$. But the dependence relation of $C$ on $m$ (see
Fig. \ref{Fig2} (b)) is more complex: (i) when $m\leq 2$ and $d\leq
6$, for the same $d$, $C$ increases with $m$; (ii) when $m\leq 2$
and $d>6$, for the same $d$, $C$ decreases with $m$; (iii) when
$m\geq 3$, for arbitrary fixed $d$, $C$ increases with $m$. The
reason for this complicated dependence relation would need further
research.

\subsection{Diameter}
The diameter of a network characterizes the maximum communication
delay in the network and is defined as the longest shortest path
between all pairs of vertices. In what follows, the notations $
\lceil x \rceil$ and $\lfloor x \rfloor$ express the integers
obtained by rounding $x$ to the nearest integers towards infinity
and minus infinity, respectively. Now we compute the diameter of
$A(d,t)$, denoted $diam(A(d,t))$ for $d\geq 2$ :


{\em Step 0}. The diameter is $1$.


{\em Steps 1 to $\lceil\frac{d}{2}\rceil$}. In this case, the
diameter is 2, since any new vertex is by construction connected to
a $(d+1)$-clique, and since any $(d+1)$-clique during those steps
contains at least $\frac{d}{2}+2$ ($d$ even) or $\frac{d+1}{2}+1$
($d$ odd) vertices from the initial $(d+2)$-clique $A(d,0)$ obtained
after step 0. Hence, any two newly added vertices $u$ and $v$ will
be connected respectively to sets $S_u$ and $S_v$, with
$S_u\subseteq V(A(d,0))$ and $S_v\subseteq V(A(d,0))$, where
$V(A(d,0))$ is the vertex set of $(A(d,0)$; however, since $\vert
S_u\vert\geq\frac{d}{2}+2$ ($d$ even) and $\vert
S_v\vert\geq\frac{d+1}{2}+1$ ($d$ odd), where $\vert S\vert$ denote
thes number of elements in set $S$, we conclude that $S_u\cap
S_v\neq{\O}$, and thus the diameter is 2.


{\em Steps $\lceil\frac{d}{2}\rceil+1$ to $d+1$}. In any of those
steps, some newly added vertices might not share a neighbor in the
original $(d+2)$-clique $A(d,0)$ obtained after step 0; however, any
newly added vertex is connected to at least one vertex of the
initial clique $A(d,0)$. Thus, the diameter equals to 3.


{\em Further steps}. Clearly, at each step $t\geq d+2$, the diameter
always lies between a pair of vertices that have just been created
at this step. We will call the newly created vertices ``outer''
vertices. At any step $t\geq d+2$, we note that an outer vertex
cannot be connected with two or more vertices that were created
during the same step $0<t'\leq t-1$. Moreover, by construction no
two vertices that were created during a given step are neighbors,
thus they cannot be part of the same $(d+1)$-clique. Thus, for any
step $t\geq d+2$, some outer vertices are connected with vertices
that appeared at pairwise different steps. Thus, there exists an
outer vertex $v_t$ created at step $t$, which is connected to
vertices $v_i's$, $1\leq i\leq t-1$, all of which are pairwise
distinct. We conclude that $v_t$ is necessarily connected to a
vertex that was created at a step $t_0\le t-d-1$. If we repeat this
argument, then we obtain an upper bound on the distance from $v_t$
to the initial clique $A(d,0)$. Let $t=\alpha (d+1)+\beta$, where
$1\leq \beta\leq d+1$. Then, we see that $v_t$ is at distance at
most $\alpha +1$ from a vertex in $A(d,0)$. Hence any two vertices
$v_t$ and $w_t$ in $A(d,t)$ lie at distance at most $2(\alpha
+1)+1$; however, depending on $\beta$, this distance can be reduced
by 1, since when $\beta\leq \lceil\frac{d}{2}\rceil$, we know that
two vertices created at step $\beta$ share at least a neighbor in
$A(d,0)$. Thus, when $1\leq p\leq \lceil\frac{d}{2}\rceil$,
$diam(A(d,t))\leq 2(\alpha +1)$, while when $\lceil\frac{d}{2}\rceil
+1\leq p\leq d+1$, $diam(A(d,t))\leq 2(\alpha +1)+1$.
One can see that these distance bounds can be reached by pairs of
outer vertices created at step $t$. More precisely, those two
vertices $v_t$ and $w_t$ share the property that they are connected
to $d$ vertices that appeared respectively at steps $t-1,t-2,\ldots
t-d-1$.

Based on the above arguments, one can easily see that for $t>d+2$,
the diameter increases by 2 every $d+1$ steps. More precisely, we
have the following result, for any $d\geq 1$ and $t\geq 1$ (when
$t=0$, the diameter is clearly equal to 1):
$$diam(A(d,t))=2(\lfloor\frac{t-1}{d+1}\rfloor +1)+f(d,t)$$
where $f(d,t)=0$ if $t-\lfloor\frac{t-1}{d+1}\rfloor (d+1)\leq
\lceil\frac{d}{2}\rceil$, and 1 otherwise.

In the limit of large $t$, $diam(A(d,t))\sim \frac{2t}{d+1}$, while
$N_t\sim [m(d+1)]^{t}$, thus the diameter is small and scales
logarithmically with the network order.

\section{Conclusion and discussion}

In summary, we have proposed and studied a network model, which is
built in an iterative fashion. At each time step, each already
existing $(d+1)$-clique, that is created at last time step, produces
$m$ new vertices. The iterative process leads to a serial of
networks with Apollonian networks corresponding to the particular
case $m=1$. We have obtained the analytical result for degree
exponent, clustering coefficient and diameter. The degree exponent
and the clustering coefficient may be adjusted to various values by
tuning the parameter $m$. Moreover, our networks consist of complete
graphs, they may represent a variety of real-life systems such as
movie actor collaboration networks, scientific collaboration
networks and networks of company directors, all of which are
composed of cliques~\cite{AlBa02,DoMe03,SaVe04,Ne03}.
\smallskip

This research was supported by the National Natural Science
Foundation of China under Grant No. 70431001.  

\end{document}